\documentclass[12pt,preprint]{aastex}

\shorttitle{Number Counts of UV Galaxies}
\shortauthors{Xu et al.}

\begin{document}

\title{Number Counts of GALEX Sources in FUV (1530{\AA}) and NUV (2310{\AA}) Bands}

\author{C. Kevin Xu \altaffilmark{1}, Jose Donas\altaffilmark{2}, 
Stephane Arnouts\altaffilmark{2}, 
Ted K. Wyder\altaffilmark{1},
Mark Seibert\altaffilmark{1},
Jorge Iglesias-P\'{a}ramo\altaffilmark{2}, 
Jeremy Blaizot\altaffilmark{2}, 
Todd Small\altaffilmark{1},
Bruno Milliard\altaffilmark{2},
David Schiminovich\altaffilmark{1},
D. Christopher Martin\altaffilmark{1},
Tom A. Barlow\altaffilmark{1},
Luciana Bianchi\altaffilmark{2},
Yong-Ik Byun\altaffilmark{3}, 
Karl Forster\altaffilmark{1},
Peter G. Friedman\altaffilmark{1},
Timothy M. Heckman\altaffilmark{5},
Patrick N. Jelinsky\altaffilmark{6},
Young-Wook Lee\altaffilmark{3},
Barry F. Madore\altaffilmark{7,8},
Roger F. Malina\altaffilmark{4},
Patrick Morrissey\altaffilmark{1},
Susan G. Neff\altaffilmark{9},
R. Michael Rich\altaffilmark{10},
Oswald H. W. Siegmund\altaffilmark{6},
Alex S. Szalay\altaffilmark{5}, and
Barry Y. Welsh\altaffilmark{6}}

\altaffiltext{1}{California Institute of Technology, MC 405-47, 1200 East
California Boulevard, Pasadena, CA 91125}
\altaffiltext{2}{Center for Astrophysical Sciences, The Johns Hopkins
University, 3400 N. Charles St., Baltimore, MD 21218}
\altaffiltext{3}{Center for Space Astrophysics, Yonsei University, Seoul
120-749, Korea}
\altaffiltext{4}{Laboratoire d'Astrophysique de Marseille, BP 8, Traverse
du Siphon, 13376 Marseille Cedex 12, France}
\altaffiltext{5}{Department of Physics and Astronomy, The Johns Hopkins
University, Homewood Campus, Baltimore, MD 21218}
\altaffiltext{6}{Space Sciences Laboratory, University of California at
Berkeley, 601 Campbell Hall, Berkeley, CA 94720}
\altaffiltext{7}{Observatories of the Carnegie Institution of Washington,
813 Santa Barbara St., Pasadena, CA 91101}
\altaffiltext{8}{NASA/IPAC Extragalactic Database, California Institute
of Technology, Mail Code 100-22, 770 S. Wilson Ave., Pasadena, CA 91125}
\altaffiltext{9}{Laboratory for Astronomy and Solar Physics, NASA Goddard
Space Flight Center, Greenbelt, MD 20771}
\altaffiltext{10}{Department of Physics and Astronomy, University of
California, Los Angeles, CA 90095}

\begin{abstract}
Number Counts of galaxies in two GALEX bands (FUV: 1530{\AA} and NUV:
2310{\AA}, both in AB magnitudes) 
are reported.  They provide for the first time in the
literature homogeneously calibrated number counts of UV galaxies
covering continuously a very wide range of UV magnitude (14 --
23.8). Both the FUV and NUV counts are inconsistent with 
a non-evolution model, while they are in good agreement with evolution
models (essentially luminosity evolution) derived from 
the high-z UV luminosity functions of Arnouts et al. (2004).
It is found that the contribution from 
galaxies detected by GALEX to the UV background is 
0.68$\pm 0.10$ nW m$^{-2}$ sr$^{-1}$ at 1530{\AA}
and 0.99$\pm 0.15$ nW m$^{-2}$ sr$^{-1}$ at 2310{\AA}.
These are 66$\pm 9$\% and 44$\pm 6$\% of the
total contributions of galaxies to the the UV background at 1530{\AA}
(1.03$\pm 0.15$  nW m$^{-2}$ sr$^{-1}$) and at 2310{\AA}
(2.25$\pm 0.32$ nW m$^{-2}$ sr$^{-1}$), respectively,
as estimated using the evolution models.
Galaxy counts and star counts in 7 regions, each contains a few
deg$^2$ GALEX coverage in an area of up to $\sim 30$ deg$^2$, are
compared with each other to study the region-by-region variance. This
shows that for the galaxy counts
the cosmic variance is comparable to the net error due
to other uncertainties. The star counts increase with decreasing
absolute Galactic latitude $|$b$|$.
\end{abstract}


\section{Introduction}
The number counts of UV galaxies as a function of the UV magnitude
provide direct measures of the density and the evolution of star
forming galaxies.  They also give strong constraints on the brightness of
the cosmic UV background. In the literature, the UV number counts have
only been measured in a few small sky areas and in some discontinued
UV magnitude ranges by several inhomogeneously calibrated UV surveys
(Milliard et al. 1992; Deharveng et al. 1994; Gardner et al. 2000;
Sasseen et al. 2002; Iglesias-P\'{a}ramo et al. 2004).  The Galaxy Evolution
Explorer (GALEX) has opened a new era of extra-galactic UV
astronomy. It is surveying the UV sky with unprecedented efficiency
and very high sensitivity. With a pyramid-like survey structure
(Martin et al 2004), GALEX presents a uniform and complete picture of
galaxy evolution in the UV domain since z$\sim 1$. In this paper, we carried
out galaxy number counts in the two GALEX bands: the FUV
(1530{\AA}) and the NUV (2310{\AA}). We have the following science goals: (1)
constraining the evolution models of UV galaxies; (2)
evaluating the cosmic variance of UV galaxies; (3) estimating
the contribution from galaxies to the cosmic background
radiation in the two UV bands; (4) providing the calibrations for the
UV luminosity functions.
\section{Data}
\begin{figure*}
\plottwo{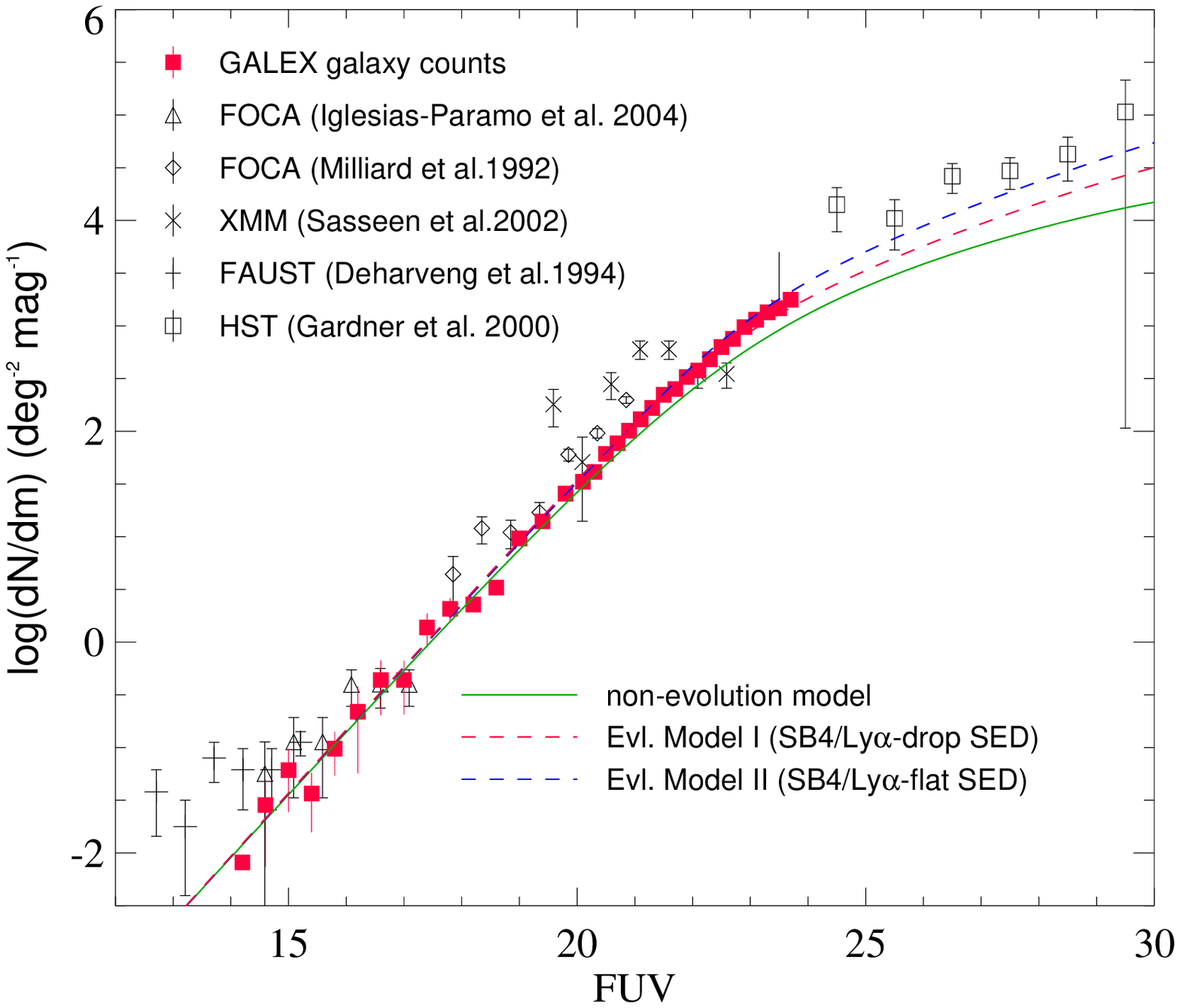}{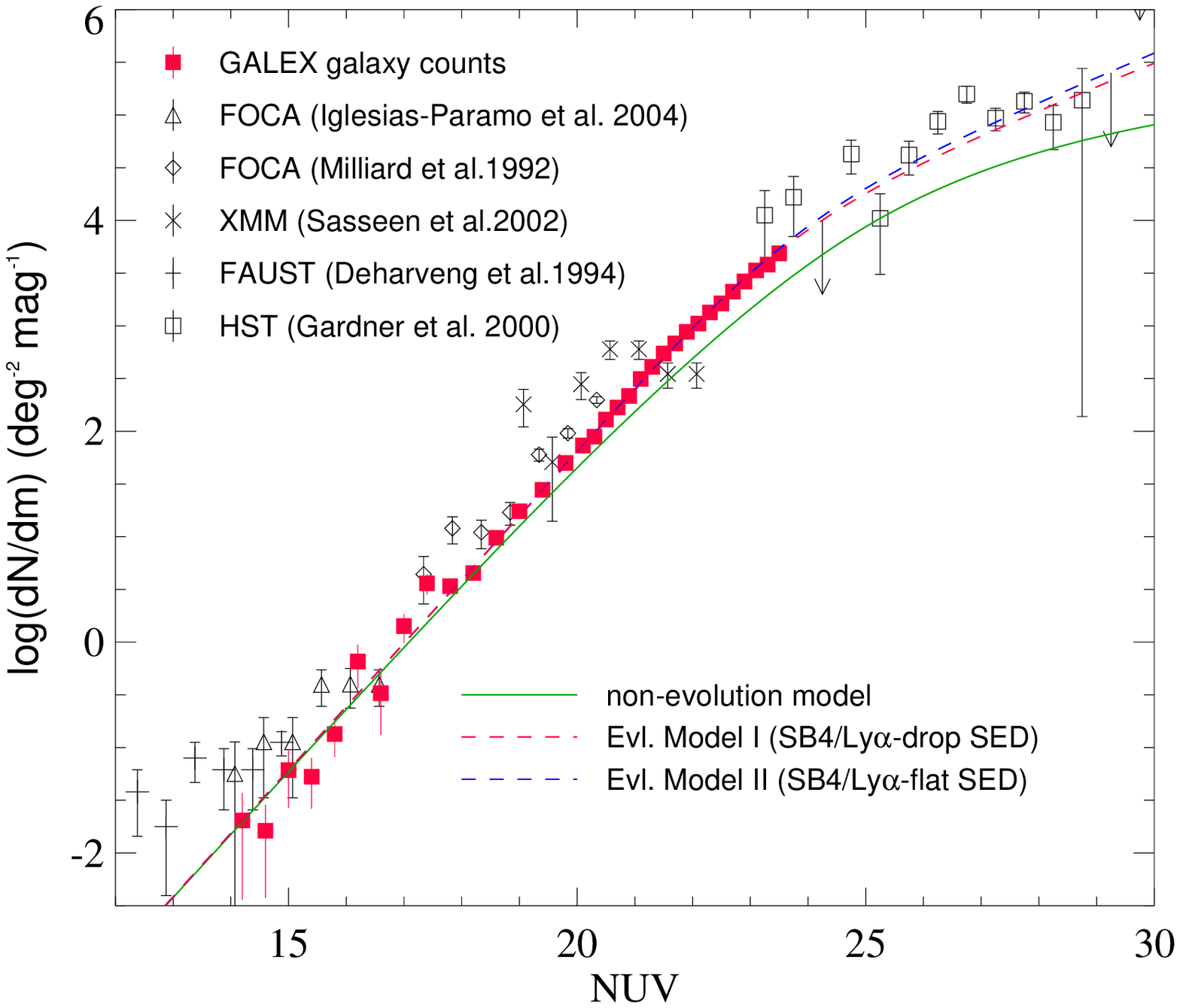}\\
\vskip-0.8cm
\caption{{\bf Left}: FUV (1530{\AA}) number counts of galaxies.
{\bf Right}: NUV (2310{\AA}) number counts of galaxies. Number 
counts in other UV bands are converted to counts in
the GALEX bands according to an assumed UV slope of $\beta=-0.8$.}
\label{fig1}
\vskip-0.5cm
\end{figure*}
Three data sets taken from the GALEX database were analyzed in this
work, yielding 17174 FUV galaxies and 41512 NUV galaxies in
the final samples for the counts (Table~1).
All the magnitudes are in the AB system. The calibrations
(IR0.2 calibration) have errors of $\sim 10\%$ 
for both FUV and NUV (Morrissey et al. 2004). 
All of the magnitudes are corrected
for the Galactic extinction using the Schlegel et al. (1998)
reddening map and the Galactic extinction curve of Cardelli et
al. (1989). 

The main data set includes 36 Medium-depth Survey (MIS) fields
taken from the GALEX internal data release IR0.2. 
The exposure time of these fields is in the 
range of 1000 -- 1700 seconds with the median of $\sim 1500$ seconds.
They all have full coverage in
the Sloan Digitized Sky Survey (SDSS hereafter) DR1 database
(Abazajian et al. 2003).  In
order to minimize the artifacts which concentrate near the
periphery of the field of view (Morrissey et al. 2004), we include only
sources within 0.45 deg radius from the field center, corresponding to
a sky coverage of 0.64 deg$^2$ for each field.  
The source lists are products of the GALEX pipeline, 
which uses a combination of the programs SExtractor
(Bertin \& Arnouts 1996) and Poissonbg, a new program written for
GALEX data, to detect and obtain photometry for sources in GALEX
images.  Poissonbg determines a background map for each GALEX image with
a method very similar to that used in SExtractor itself except that
the background is estimated using the mean and counts due to sources
are iteratively clipped out using the full Poisson distribution,
rather than assuming Gaussian statistics which is not appropriate
for low counts rate found commonly in GALEX images.
Poissonbg also creates a detection threshold map for a
user-specified probability using the background image and
the full Poisson distribution.  The original image is divided by the
detection threshold map and given as the detection image to SExtractor
while the SExtractor measurements are done on the
background-subtracted image. The GALEX magnitude is taken from the
MAG$_{\rm AUTO}$ of the SExtractor. Key SExtractor/Poissonbg
parameters used by the Pipeline are listed in Table A1
(electronic table). The second data set
contains 3 Deep Survey (DIS) fields (T$_{exp}=$25536s, 23591s, 5607s), 
all in the Spitzer First Look
Survey (FLS) area.  These fields are also covered by SDSS DR1. 
In these DIS fields the source confusion becomes significant. Hence fine
tunings in the SExtractor/Poissonbg 
parameters have been applied (see electronic Table A2).
A slight dimming (on the order of 0.1 -- 0.2 mag) of the
fluxes at magnitudes fainter than 23 was found in the later stage
of data analysis, resulting from a small ($\sim 2.5\%$)
positive bias of our present sky background estimator. This was corrected
empirically according to results of the artificial source
simulations (see Section 3).
The minimum
distance among the 3 DIS fields is 0.86 deg. In order to avoid source
duplication, sources were included only when they are within 0.43 deg
radius from the field center, corresponding to a coverage of 0.58
deg$^2$ per DIS field.  The third data set consists of 95
bright GALEX galaxies (NUV $< 16$) selected by Buat et
al. (2004) in the study of the FIR emission of UV galaxies. The
characterization of these galaxies can be found in Buat et al. (2004).  

Matches with the SDSS catalogs were carried out for the MIS and DIS
sources with a search radius of $4''$. Less than 3\% of
FUV and NUV sources brighter than 22.5 mag (extinction corrected) 
have no SDSS counterparts. These are predominantly spurious sources
due to artifacts such as pieces of shredded bright extended sources.
However, optical counterparts of $r>22$ of fainter GALEX sources 
may be missed by SDSS.
In the magnitude range of 22.5 and 23.8, about 10\%
and 20\% of FUV and NUV sources do not have SDSS counterparts,
respectively. Detailed inspections of the GALEX images 
showed that most of these sources
are real.  The SDSS star/galaxy classifications were adopted for MIS/DIS
sources with SDSS counterparts. Sources brighter
than 22.5 mag and without SDSS counterparts are deemed spurious and
dropped off from the sample. Sources with UV magnitudes
in the range of 22.5 and 23.8 mag and without SDSS 
counterparts were included as galaxies. 
\section{Bias Corrections}
\noindent{\bf Incompleteness}:
We adopt the Monte Carlo simulation algorithm developed by Smail et
al. (1995) in assessing the incompleteness of GALEX
catalogs. Artificial sources of a given UV magnitude were added
randomly to a GALEX image. Then the same source extractor which
produced the original catalog from the same image was applied and the
results were checked for non-detections of the added sources. This
gives an estimate of the incompleteness.  In order to
reproduce the real point spread function (PSF) and the photon noise,
the artificial sources were created by dimming the bright sources
extracted from the same image, and the photon counts in the region
where an artificial source is added were randomized according to the
Poisson probability function. The counts in each field
were truncated at the magnitude where the incompleteness is becoming
larger than 20\%. In magnitude bins brighter than the completeness
cut-off, the error due to the incompleteness correction is estimated
conservatively to be 50\% of the correction.

\noindent{\bf Spurious Sources}:
The spurious source fraction as function of the
non-dereddened apparent magnitude was
estimated by counting single-detection sources in
regions with multipule coverages ('repeatability' method). 
The effect of incompleteness was accounted for in the analysis. 
The error is estimated to be 50\% of the correction.
\begin{deluxetable}{llllllllllll}
\tablecaption{FUV and NUV Number Counts of Galaxies\label{lum}}
\tablewidth{0pt}
\tablehead{
 \colhead{FUV} & \colhead{counts} & \colhead{f\tablenotemark{a}} 
& \colhead{$g_{gal}$} & \colhead{N} & \colhead{area} &
 \colhead{NUV} & \colhead{counts} & \colhead{f\tablenotemark{a}} 
& \colhead{$g_{gal}$} & \colhead{N} & \colhead{area} \\
 \colhead{mag} & \colhead{deg$^{-2}$mag$^{-1}$} & \colhead{}
& \colhead{}& \colhead{}& \colhead{deg$^2$} &
 \colhead{mag} & \colhead{deg$^{-2}$mag$^{-1}$} & \colhead{}
& \colhead{ }& \colhead{ }& \colhead{deg$^2$}
}
\startdata
  14.2 &   0.008 $\pm$   0.008 & 1.00 & 1.00  &    2 &615.00 &  14.2 &   0.020 $\pm$  0.015 & 1.00 & 1.00 &   5 &615.00 \\
  14.6 &   0.029 $\pm$   0.018 & 1.00 & 1.00  &    7 &615.00 &  14.6 &   0.016 $\pm$  0.012 & 1.00 & 1.00 &   4 &615.00 \\
\enddata
\tablecomments{Table 1 is published in its entirety in the electronic edition
of The Astrophysical Journal Letters. A portion is shown here for guidance
regarding its form and content.}
\tablenotetext{a}{$f=(1-f_{spur})/(1-f_{incomp})$, where $f_{spur}$ is the 
correction for the spurious sources, and $f_{incomp}$ the correction
for the incompleteness. 
}
\end{deluxetable}

\noindent{\bf Star/Galaxy Classification Errors and Contamination of AGNs}:
The biases due to mis-classifications of
sources by SDSS and due to
AGN contamination are accessed using the results of photo-z processing
of GALEX sources in the MIS fields. By fitting
fluxes in 7 bands (5 SDSS bands plus 2 GALEX bands) with templates of
stars, galaxies and AGNs, the photo-z processing makes the best use of
the information in the SEDs. Although there are still substantial
uncertainties for individual sources, the
statistics derived from these results are robust. 
Galaxy counts in a given magnitude bin were derived using the formula
\begin{equation}
c_{gal} = c_{ext} \times g_{ext} + c_{pnt} \times g_{pnt}
= c_{ext} \times g_{gal}
\end{equation}
where  $c_{ext}$ is the number counts of SDSS extended sources,
$g_{ext}$ the fraction of galaxies among SDSS extended sources,
$c_{pnt}$ the number counts of SDSS point sources, $g_{pnt}$ 
the fraction of galaxies among these sources, 
and $g_{gal} = g_{ext} + g_{pnt}\times c_{pnt}/c_{ext}$.
Both $g_{ext}$
and $g_{pnt}$ were taken from the photo-z results. 
A 5\% error was
assigned to the MIS and DIS counts in all FUV and NUV magnitude bins
to take into account the uncertainty due to mis-classifications.

\noindent{\bf Errors in Foreground Extinction Correction}:
The following errors can occur to the extinction correction:
(1) incorrect UV extinction law in average; (2) region-to-region
variations in the UV extinction law; (3) insufficient angular 
resolution of the Schlegel map. Based on results of an extensive
test involving large samples of low and high extinction regions,
a conservative 10 percent extinction correction 
error was assigned to the counts.

\noindent{\bf Eddington Bias}:
The Eddington bias (Eddington 1913) is corrected
using the algorithm developed by Hogg \& Turner (1998, Eq.(4)), assuming the
slope of the counts near the detection limit to be $p=1$. 
%
%
%
\begin{figure*}
\vskip-1.0cm
\plotone{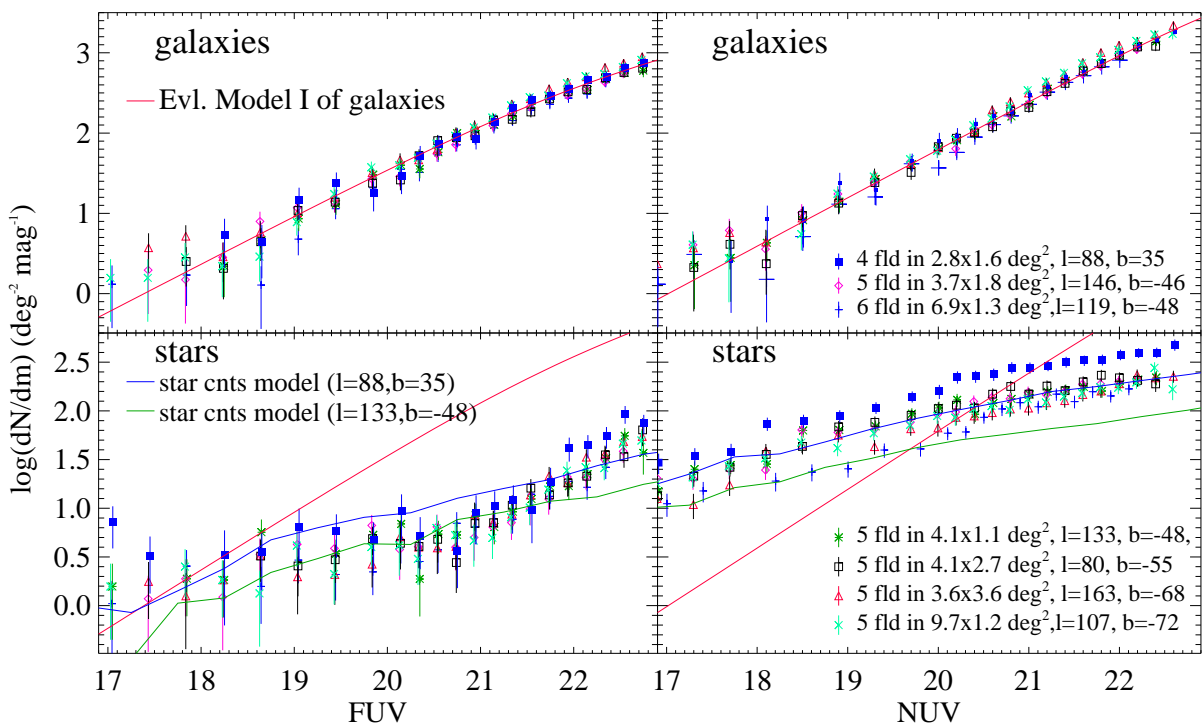}
\vskip-0.8cm
\caption{FUV and NUV number counts of 
galaxies and stars in 7 regions with MIS and DIS coverage.
}
\label{fig2}
\vskip-0.8cm
\end{figure*}
%
\section{Results}
The number counts of galaxies in the FUV and NUV bands (Table~1) 
cover continuously the magnitude
range of 14.0 -- 23.8 for FUV and 14.0 -- 23.6 for NUV.  
The error bars include not only
Poisson errors, errors due to incompleteness and spurious sources
corrections, and errors due to source classifications and
extinction correction, but
also errors due to cosmic variance
calculated using the formulism of Glazebrook et al. (1994),
which are in general on the same order of
Poisson errors. Note that the cosmic variance
might be underestimated by the assumption that individual GALEX 
fields are independent with each other.
In Fig.1 the results are compared with
UV counts taken from the literature, and with predictions
of evolution models. Both the FUV counts and the NUV 
counts are slightly lower than the FAUST and FOCA counts
(Deharveng et al. 1994; Milliard et al. 1992; Iglesias-P\'{a}ramo et al. 2004)
in bins brighter than 21 mag. This is likely to be 
due to a calibration difference between GALEX
and FAUST/FOCA (see also Wyder et al. 2004). 
It should also be noted that counts of bright
UV galaxies by  Deharveng et al. (1994) and
Iglesias-P\'{a}ramo et al. (2004) are biased to higher
values by galaxy clusters.
The XMM-Newton OM-UVW2 counts (Sasseen et al. 2002) appear to be
incomplete in bins fainter than 21 mag,
where they decrease with increasing magnitude.
The GALEX counts are consistent with the
HST counts (Gardner et al. 2000) in the magnitude bins 
where the two overlap.

Both the FUV and NUV counts
are inconsistent with predictions by a non-evolution model,
which was constructed using the local luminosity 
functions in the FUV and NUV (Wyder et al. 2004). 
The K-correction was estimated using an SED based on the 
SB4 class spectrum of Kinney et al. (1996) 
at $\lambda > 1200${\AA}, a dropping spectrum
(by a factor of $\sim 2$) between 1200 --- 1000{\AA},
and a sharp cut-off shortward of
912{\AA}. The SB4 spectrum has $\beta = -0.8$.
This is close to the value calculated from the ratio of the FUV
and NUV luminosity densities in the local universe by 
Wyder et al. (2004), after allowing for the difference in 
the calibration
zero points (0.03 mag for FUV and -0.10 mag for NUV, Morrissey et al. 2004) 
used in these two works.
The high redshift UV luminosity functions of 
Arnouts et al. (2004), which indicate a UV luminosity density
evolution of $(1+z)^{2.5\pm 0.7}$ to $z \sim 1$ (Schiminovich et al. 2004),
can be best fitted by an analytic model
with L$_* \propto (1+z)^{2.5}$,
$\phi_* \propto (1+z)^{-1.3}$ and $\alpha \propto (1+z)^{-1.3}$.
In the expression for the
integrated luminosity density of different redshift,
${\rm \rho_L (z)=\phi_* (z) L_* (z) \Gamma (2 - \alpha (z))}$,
the evolution of $\phi_*$ and that of $\alpha$ cancel with
each other for $0 < \alpha < 2$, leading to ${\rm \rho_L(z) \sim
L_* (z)}$. Therefore this is essentially a luminosity 
evolution model.
Predictions by two models, both assuming the same evolution parameters
as given above but using different UV SEDs, are plotted 
in Fig.1: Model I 
assumes the same SED as that of the no evolution model, 
and the SED assumed in Model II 
is otherwise the same except for a flat spectrum
between 1200 --- 1000{\AA}.
These two models should embrace the uncertainties in the
SED of wavelengths shorter than the Ly$\alpha$ due both
to the Ly$\alpha$ absorption and dust extinction (e.g. Buat et al. 2002), 
which affects significantly the K-corrections in GALEX bands.
In the bins covered by the GALEX counts,
the two evolution models give almost identical results,
both fit the data very well. In 
fainter bins, the two models show significant difference
in the FUV band due to different K-corrections. Both models
fit the HST NUV counts well, given
the substantial uncertainties of the data,
but are slightly lower than the HST FUV counts.

Using the mean of the integral counts predicted by  
Model I and Model II, and the beam size of the DIS FUV
maps (FWHM=$5.4''$) and that of DIS NUV maps ($5.6''$),
we estimate that the 40-beam-per-source confusion limits of the FUV and
NUV maps are FUV=25.3 and NUV=24.0, respectively.
The major contribution to the incompleteness at the
boundary of the NUV counts (NUV=23.6) is from the source confusion,
indicating the NUV counts being confusion limited. The confusion limit
of the DIS FUV maps, on the other hand, is significantly deeper than 
the last bin of the FUV counts. The contribution from
galaxies detected by GALEX to the UV background is
0.68$\pm 0.10$ nW m$^{-2}$ sr$^{-1}$ at 1530{\AA}
and 0.99$\pm 0.15$ nW m$^{-2}$ sr$^{-1}$ at 2310{\AA}.
Extrapolating the counts using the above two evolution models
and integrating down to flux=0, the total extragalactic
UV background due to galaxies is 1.03$\pm 0.15$ 
nW m$^{-2}$ sr$^{-1}$ at 1530{\AA}
and 2.25$\pm 0.32$ nW m$^{-2}$ sr$^{-1}$ at 2310{\AA}, respectively.
The errors include both model uncertainties and GALEX calibration
uncertainties. Our results are significantly
lower than the results of Gardner et al. (2000) who found that
the contributions from resolved sources to the UV background are
3.9$^{+1.1}_{-0.8}$ nW m$^{-2}$ sr$^{-1}$ and
3.6$^{+0.7}_{-0.5}$ nW m$^{-2}$ sr$^{-1}$ at 1595{\AA} and 2365{\AA},
respectively. 
Our results are in good agreement with that of Armand et al. (1994),
who found that the contribution from galaxies to the UV
background at 2000{\AA} is 40 -- 140 photons cm$^{-2}$ $s^{-1}$ 
{\AA}$^{-1}$ sr$^{-1}$, corresponding to 0.80 -- 2.78 nW m$^{-2}$ sr$^{-1}$.

In Fig.2 FUV and NUV number counts of galaxies and stars in 7 regions
with MIS and DIS coverage are presented in 4 panels. For galaxy
counts the error bars do not include the cosmic variance in these
plots. It appears that differences between galaxy counts in different
regions are consistent with the error bars, indicating that the cosmic
variance is not very significant for UV surveys covering more than
a few square degrees.
The counts of stars show a significant trend in the sense that the
less the absolute Galactic latitude ($|$b$|$), the higher the
counts. The NUV counts of stars are rather flat, higher than galaxy
counts in bins brighter than $\sim 21$ mag.  The FUV counts of stars
are lower than the counts of galaxies in bins fainter than $\sim 18$
mag.  These data are compared with predictions of a star counts model.
The model closely follows the Bahcall-Soneira model (Bahcall 1986)
with an updated parameterization of the Galactic spheroid from Gould
et al.  (1998). The stellar luminosity function of Wielen et al. (1983)
is used. An additional disk white dwarfs population (scale height of
275 kpc) is included, with a luminosity function drawn from Liebert et
al. (1988).
The model generally reproduces the trends of the observed
star counts, though the predicted counts in both bands
are slightly flatter than the observations.

\vskip0.2truecm
\noindent{\it Acknowledgments}:
GALEX (Galaxy Evolution Explorer) is a NASA Small Explorer, launched
in April 2003.  We gratefully acknowledge NASA's support for
construction, operation, and science analysis for the GALEX mission,
developed in cooperation with the Centre National d'Etudes Spatiales
of France and the Korean Ministry of Science and Technology.


\end{document}